\documentclass[12pt]{article}
\usepackage[centertags]{amsmath}
\usepackage{amsfonts}
\usepackage{epsfig}
\usepackage{amssymb}
\usepackage{amsthm}
\usepackage{newlfont}
\usepackage{amstext}

\theoremstyle{plain}
\newtheorem{Th}{Theorem}[section]
\newtheorem{Cor}[Th]{Corollary}

\newtheorem{Prop}[Th]{Proposition}
\theoremstyle{definition}
\newtheorem{Def}{Definition}[section]

\theoremstyle{remark}

\numberwithin{equation}{section}
\newcommand{\NN}{{\mathbb N}}
\newcommand{\PP}{{\mathbb P}}
\newcommand{\RR}{{\mathbb R}}

\newcommand{\ZZ}{{\mathbb Z}}

\newcommand{\cL}{{\cal L}}

\newcommand{\bx}{{\boldsymbol x}}

\newcommand{\bX}{{\boldsymbol X}}

\newcommand{\tbX}{\tilde{\bX}}

\newcommand{\tH}{\tilde{H}}
\newcommand{\tQ}{\tilde{Q}}

\newcommand{\D}{{\Delta}}
\newcommand{\tD}{\tilde{\Delta}}

\begin{document}

\title{Lattice geometry of the Hirota equation}
\author{Adam Doliwa \\
         Instytut Fizyki Teoretycznej, Uniwersytet Warszawski\\
         ul. Ho{\.z}a 69, 00-681 Warszawa, Poland\\
         e-mail: {\tt doliwa@fuw.edu.pl}}
\date{ }
\maketitle

\begin{abstract}
Geometric interpretation of the Hirota equation is presented as
equation describing the Laplace sequence of two-dimensional
quadrilateral lattices. Different forms of the equation are given
together with their geometric interpretation: (i)~the discrete
coupled Volterra system for the coefficients of the Laplace
equation, (ii)~the gauge invariant form of the Hirota equation for
projective invariants of the Laplace sequence, (iii)~the discrete
Toda system for the rotation coefficients and (iv)~the original
form of the Hirota equation for the $\tau$-function of the
quadrilateral lattice.

\medskip


\end{abstract}

\section{Introduction}

The integrable discrete geometry deals with lattice submanifolds
described by integrable equations. Among the integrable discrete
(difference) equations an important role is played by the Hirota
equation~\cite{Hirota} which is the discrete analog of the
two-dimensional Toda system~\cite{Mikhailov}. Both the Toda system
and the Hirota equation are important in the theory of integrable
equations and in their applications. It turns out that the 
two-dimensional Toda system was studied in classical differential
geometry and describes the so called Laplace sequences of
two-dimensional conjugate nets~\cite{DarbouxIV}.

The lattice geometric interpretation of the discrete Toda system
was found in~\cite{DCN} and is based on the observation that the
discrete analog of the conjugate net on a surface, which is given by
the two-dimensional lattice made of planar
quadrilaterals~\cite{Sauer}, allows for definition of the
corresponding Laplace sequence of discrete conjugate nets. Since
then, the multidimensional quadrilateral lattice~\cite{MQL} 
(multidimensional 
lattice made of planar quadrilaterals --
the {\it integrable}
discrete analog of a multidimensional conjugate net) became
one of the central notions of the integrable discrete geometry. 
In particular, many
classical results of the theory of conjugate nets and of their reductions
have been generalized recently to the 
discrete level \cite{BP2,MDS,CDS,DMS,TQL,KoSchief2,q-red,DS-sym}.

The goal of the present article is to reinterpret and expand
results obtained in~\cite{DCN} using new notions provided by the
general theory of quadrilateral lattices. Different formulations
of the Hirota equations are reviewed and, for each of them, the
geometric interpretation of the corresponding functions is given.

In Sections~\ref{sec:A} and~\ref{sec:K} we reformulate, using more 
convenient notation, results
found in~\cite{DCN}. In
Section~\ref{sec:Q} we present the discrete analog of the standard
version of the Toda system as an equation
governing Laplace transformations of the rotation
coefficients. Then in Section~\ref{sec:tau}, basing on the
geometric interpretation of the $\tau$--function of the
quadrilateral lattice given in~\cite{DS-sym}, we show that the
$\tau$--functions of the Laplace sequence of quadrilateral
lattices solve the original Hirota's form of the discrete Toda
system --- this last result fills out the missing point of 
paper~\cite{DCN}.

\section{Laplace transformations of quadrilateral lattices}
\label{sec:A}
\begin{Def}
Two dimensional quadrilateral lattice is a mapping of the two
dimensional integer lattice into $M$ dimensional projective space
such that elementary quadrilaterals of the lattice are planar:
\begin{equation*}
x:\ZZ^2 \rightarrow \PP^M, \qquad T_1T_2 x \in \langle x, T_1 x ,
T_2 x \rangle .
\end{equation*}
\end{Def}
In the above definition $T_i$, $i=1,2$, denotes the shift operator
along the $i$-th direction of the lattice.

In the non-homogenous coordinates of the projective space a two
dimensional quadrilateral lattice is represented by the mapping
$\bx : \ZZ^2 \rightarrow \RR^M$ satisfying the Laplace
equation~\cite{DCN}
\begin{equation}  \label{eq:Laplace}
\D_1\D_2\bx=(T_{1} A_{12})\D_2\bx+ (T_2 A_{21})\D_2\bx,
\end{equation}
which is equivalent to the planarity condition;
here $\D_i=T_i - 1$, $i=1,2$, is the partial difference operator,
and the functions $A_{12},A_{21}:\ZZ^2 \rightarrow \RR$, define the
position of the point $T_1 T_2 \bx$ with respect to the points
$\bx$, $T_1\bx$ and $T_2 \bx$.

\begin{figure}
\begin{center}
\epsffile{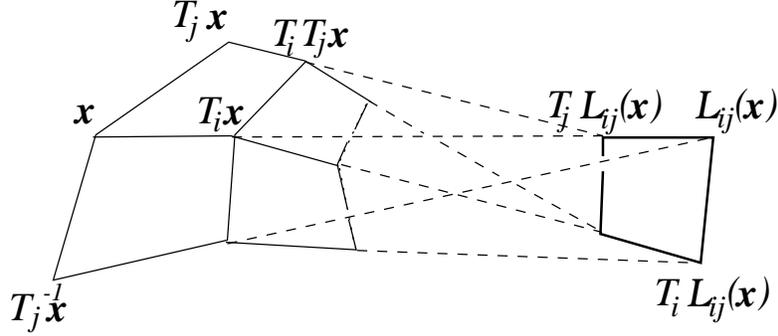}
\end{center}
\caption{Laplace transformation of quadrilateral lattices}
\label{fig:Lijx}
\end{figure}

Planarity of elementary quadrilaterals 
implies that the "tangent" lines containing opposite sides of an
elementary
quadrilateral intersect. The Laplace transformation $\cL_{ij}$,
$i\ne j$ of the lattice $x$ is defined \cite{DCN,TQL} as
intersection of the line $\langle x, T_i x \rangle$ with the line
$\langle T_j^{-1}x, T_iT_j^{-1} x \rangle$ (see Figure~\ref{fig:Lijx}). 

Using elementary
calculations one can show the following results~\cite{DCN,TQL}.
\begin{Prop}
In the non-homogenous representation the Laplace transformation
$\cL_{ij}(\bx)$ of the quadrilateral lattice $\bx$ is given by
\begin{equation} \label{eq:Lij-x}
\cL_{ij}(\bx) = \bx - \frac{1}{A_{ji}}\D_i\bx .
\end{equation}
\end{Prop}
\begin{Prop}
The Laplace transformations of quadrilateral lattices are again
quadrilateral lattices and the corresponding transformations of
the coefficients of the Laplace equation~\eqref{eq:Laplace} of the
lattice $\cL_{ij}(\bx)$ read
\begin{align} \label{eq:LijAij}
\cL_{ij}(A_{ij}) &= \frac{A_{ji}}{T_jA_{ji}} (T_iA_{ij} +1) -1 ,
\\ \label{eq:LijAji} \cL_{ij}(A_{ji})  & =  T_j^{-1}\left(
\frac{T_i\cL_{ij}(A_{ij})}{\cL_{ij}(A_{ij})} \left( A_{ji} +
1\right)\right) -1 .
\end{align}
\end{Prop}
\begin{Prop} \label{prop:Lij-Lji}
Under the assumption that the transformed lattices are
non-degenerate, i.e. their quadrilaterals do not degenerate to
segments or points, we have
\begin{equation}
\cL_{ij} \circ \cL_{ji} = \cL_{ji} \circ \cL_{ij} = \text{id.}
\end{equation}
\end{Prop}
In this way given two dimensional quadrilateral lattice $\bx$ one
can define a sequence of quadrilateral lattices
\begin{equation*}
\bx^{(l)} = \cL_{12}^l(\bx), \quad l\in \NN , \quad \cL_{12}^{-1}
= \cL_{21} .
\end{equation*}
In analogy to the Laplace sequence of conjugate nets, the above
sequence can be called the Laplace sequence of quadrilateral
lattices. Equations~\eqref{eq:LijAij}-\eqref{eq:LijAji} can be
then rewritten in the form
\begin{align}
\frac{\D_2A_{21}^{(l)}}{A_{21}^{(l)}} & = \frac{T_1A_{12}^{(l)}-
A_{12}^{(l+1)}}{(T_1A_{12}^{(l)}+1)(A_{12}^{(l+1)}+1)} \; ,
\label{eq:T3A12} \\ \frac{\D_1A_{12}^{(l)}}{A_{12}^{(l)}} & =
\frac{T_2A_{12}^{(l)}-
A_{21}^{(l-1)}}{(T_2A_{21}^{(l)}+1)(A_{21}^{(l-1)}+1)} \; ,
\label{eq:T3A21}
\end{align}
which is the discrete analog of the coupled Volterra system.

\section{Projective invariants of the Laplace sequence}
\label{sec:K}

The planarity of elementary quadrilaterals of the quadrilateral
lattice and the construction of the Laplace sequence are
essentially of the projective nature~\cite{DCN}. It would be
therefore interesting to know the pure projective-geometric
version of the equation describing the Laplace sequence of
quadrilateral lattices.

The basic numeric invariant of projective transformations is the
so called cross-ratio of four collinear points, which is given in
the affine representation as
\begin{equation*}
\text{cr}(a, b ;c, d) = \left( \frac{c-a}{c-b} \right) :\left(
\frac{d-a}{d-b} \right) ;
\end{equation*}
notice the simple identity
\begin{equation} \label{eq:cr-cr}
\text{cr}(a,b;c,d) = \text{cr}(b, a ;d ,c) .
\end{equation}
Define the function $K_{ij}$ as the cross-ratio of $x$,
$\cL_{ij}(x)$, $T_i x$ and $T_j\cL_{ij}(x)$. Elementary
calculations show that
\begin{align*}
T_i\bx - \cL_{ij}(\bx) & = \frac{1+ A_{ji}}{A_{ji}}\D_i\bx ,\\
T_j\cL_{ij}(\bx) - \bx & = -\frac{T_i A_{ij} + 1 }{ T_j A_{ji} }
\D_i \bx , \\ T_j\cL_{ij}(\bx) - \cL_{ij}(\bx) & = \left(
\frac{1}{A_{ji}} - \frac{1+ T_iA_{ij}}{T_jA_{ji}} \right) \D_i\bx,
\end{align*}
and, therefore,
\begin{equation*}
K_{ij} = \text{cr}( \bx , \cL_{ij}(\bx) ; T_i \bx ,
T_j\cL_{ij}(\bx)) = \frac{A_{ji}(T_iA_{ij}+1) - T_jA_{ji}} { (1 +
T_iA_{ij}) (1+A_{ji}) }.
\end{equation*}
Equations \eqref{eq:LijAij} and \eqref{eq:LijAji} allow to find
the Laplace transforms of the projective invariants~\cite{DCN}
\begin{align}
\cL_{ij}(K_{ij}) & = T_j^{-1}\left( \frac{K_{ij} (T_iT_j
K_{ij})}{(T_i K_{ij})(T_j K_{ij})} \frac{(T_i K_{ij} +
1)(T_jK_{ij} + 1)}{T_i K_{ji} + 1} -1 \right) \; ,
\label{eq:LijKij}
\\ \label{eq:LijKji} \cL_{ij}(K_{ji}) & = K_{ij} \; ;
\end{align}
notice that equation~\eqref{eq:LijKji} is a simple consequence of
Proposition~\ref{prop:Lij-Lji} and property~\eqref{eq:cr-cr} of
the cross-ratio.

Equations~\eqref{eq:LijKij} and \eqref{eq:LijKji} can be rewritten
in terms of the function $K=K_{12}$ in the following form
\begin{equation} \label{eq:K}
T_2 \left( \frac{K^{(l+1)}+1}{K^{(l)}+1} \right)
T_1\left(\frac{K^{(l-1)}+1}{K^{(l)}+1} \right) = \frac{(T_1T_2
K^{(l)}) K^{(l)}}{(T_1K^{(l)})(T_2K^{(l)})} \; ,
\end{equation}
known as the gauge invariant form of the Hirota equation.

\section{Rotation coefficients of the quadrilateral lattice}
\label{sec:Q} 
\begin{figure}
\begin{center}
\epsffile{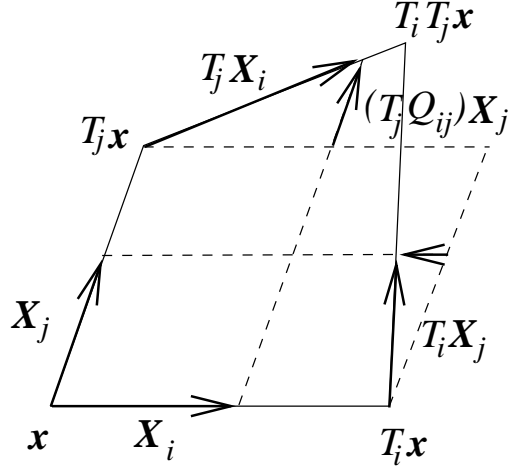}
\end{center}
\caption{Definition of the rotation coefficients}
\label{fig:forward}
\end{figure}

As it was shown in \cite{MQL} it is convenient to
reformulate the Laplace equation (\ref{eq:Laplace}) as a first
order system. We introduce the suitably scaled tangent vectors
$\bX_i$, $i=1,2$,
\begin{equation}  \label{def:HX}
\D_i\bx = (T_iH_i) \bX_i,
\end{equation}
in such a way that the $j$-th variation of $\bX_i$ is proportional
to $\bX_j$ only (see Figure~\ref{fig:forward})
\begin{equation} \label{eq:lin-X}
\D_j\bX_i = (T_j Q_{ij})\bX_j,    \; \; \; i\ne j ,
\end{equation}
the coefficients $Q_{ij}$ in equation \eqref{eq:lin-X} are called
the rotation coefficients. 

The scaling factors $H_i$ in equation
\eqref{def:HX}, called the Lam\'e coefficients, satisfy the linear
equations
\begin{equation*} \label{eq:lin-H}
\D_iH_j = (T_iH_i) Q_{ij}, \; \; \; i\ne j \;,
\end{equation*}
adjoint to \eqref{eq:lin-X};
moreover
\begin{equation*}   \label{def:A-H}
A_{ij}= \frac{\D_j H_i}{H_i} \; , \; \; i\ne j \; .
\end{equation*}
The Laplace transformation of the Lam\'e coefficients and the
scaled tangent vectors was found in \cite{TQL} and is presented in
the following
\begin{Prop}
The Lam\'e coefficients of the transformed lattice read
\begin{align*}
\cL_{ij}(H_i) &= \frac{H_j}{Q_{ij}} \; , \\ \cL_{ij}(H_j) &=
T_j^{-1}\left( Q_{ij}\D_j \left(\frac{H_j}{Q_{ij}}  \right)
\right) \; ,
\end{align*}
the tangent vectors of the new lattice are given by
\begin{align*}
\cL_{ij}(\bX_i) & = - \D_i\bX_i + \frac{\D_iQ_{ij}}{Q_{ij}} \bX_i
\; , \\ \cL_{ij}(\bX_j) & = -\frac{1}{Q_{ij}}\bX_i \; .
\end{align*}
\end{Prop}
From above formulas  follow transformation rules for the
rotation coefficients.
\begin{Prop}
The rotation coefficients transform according to
\begin{align} \label{eq:Lij-Qij}
\cL_{ij}(Q_{ij}) &= T_j^{-1}\left( T_iQ_{ij} -\frac{ Q_{ij}\:
T_iT_jQ_{ij}}{T_jQ_{ij}} \left( 1-(T_iQ_{ji})(T_jQ_{ij}) \right)
\right),
\\ \label{eq:Lij-Qji} \cL_{ij}(Q_{ji})&= \frac{1}{Q_{ij}} .
\end{align}
\end{Prop}
Equations \eqref{eq:Lij-Qij} and \eqref{eq:Lij-Qji} can be
rewritten in terms of the function $Q=Q_{12}$ as
\begin{equation} \label{eq:Q}
\D_2\frac{\D_1 Q^{(l)}}{Q^{(l)}}= T_1\left(
\frac{T_2Q^{(l)}}{Q^{(l-1)}} \right) - \frac{T_2
Q^{(l+1)}}{Q^{(l)}} .
\end{equation}

\section{Geometry of the $\tau$--function}
\label{sec:tau} 
\begin{figure}
\begin{center}
\epsffile{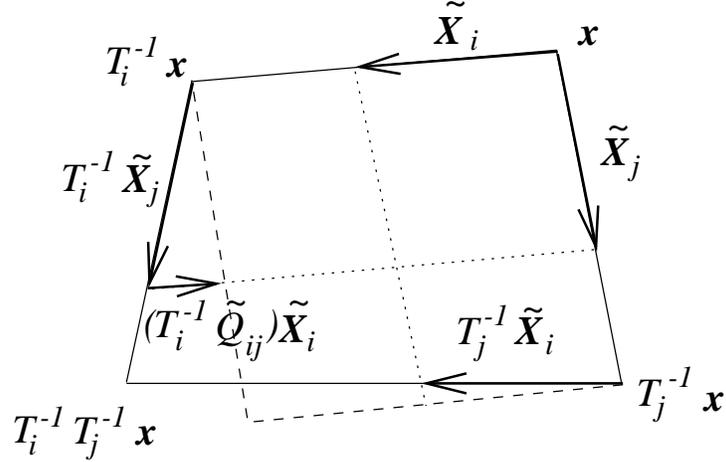}
\end{center}
\caption{Definition of the backward data}
\label{fig:back}
\end{figure}

To give the geometric meaning to the
$\tau$-function let us introduce~\cite{DS-sym} the vectors
$\tbX_i$ pointing in the negative directions 
\begin{equation*} \label{eq:b-H-X}
\tD_i\bx = (T_i^{-1}\tH_i ) \tbX_i \; , \quad \text{or} \quad
\D_i\bx = \tH_i (T_i\tbX_i)  \, ,
\end{equation*}
where $\tD_i= T_i^{-1} - 1$ is the backward difference operator.
The scaling factors $\tH_i$ (the backward Lam\'e coefficients)
are chosen in such a way that the
$\tD_i$ variation of $\tbX_j$ is proportional to $\tbX_i$ only
(see Figure~\ref{fig:back}). We
define the backward rotation coefficients $\tQ_{ij}$ as the
corresponding proportionality factors
\begin{equation} \label{eq:lin-bX}
\tD_i\tbX_j = (T_i^{-1} \tQ_{ij})\tbX_i \; , \qquad \text{or}
\quad \D_i\tbX_j =  (T_i\tbX_i)\tQ_{ij},    \quad i\ne j \; .
\end{equation}

The backward Lam\'e coefficients $\tH_i$ satisfy then the following 
system of
linear equations
\begin{equation*} \label{eq:lin-bH}
\D_j\tH_i =  (T_j\tQ_{ij})\tH_{j},   \; \; \; i\ne j \; ,
\end{equation*}
adjoint to system~\eqref{eq:lin-bX}

Since the forward and backward rotation coefficients 
$Q_{ij}$ and $\tQ_{ij}$
describe the same lattice $\bx$ but from different points of view,
then one cannot expect that they are independent. Indeed, defining
the functions $\rho_i:\ZZ^2\to \RR$, $i=1,2$, as the
proportionality factors between $\bX_i$ and $T_i\tbX_i$ (both
vectors are proportional to $\D_i\bx$):
\begin{equation*} \label{eq:def-rho}
\bX_i = - \rho_i ( T_i\tbX_i) \; , \qquad T_iH_i = -
\frac{1}{\rho_i}\tH_i \; , \quad i=1,2 \; ,
\end{equation*}
we have the following result~\cite{DS-sym}

\begin{Prop} \label{prop:bf-data}
The forward and backward rotation coefficients 
of the lattice $\bx$ are related
through the following formulas
\begin{equation*} \label{eq:Q-Qt}
\rho_j T_j\tQ_{ij} =  \rho_i T_iQ_{ji} \; ,
\end{equation*}
and the factors $\rho_i$ are first potentials satisfying equations
\begin{equation} \label{eq:rho-constr}
\frac{T_j\rho_i}{\rho_i} = 1 - (T_iQ_{ji})(T_jQ_{ij}) \; , i\ne j
\; .
\end{equation}
\end{Prop}

The right hand side of equation~(\ref{eq:rho-constr}) is symmetric with
respect to the interchange of $i$ and $j$, which implies the
existence of a potential $\tau:\ZZ^2\to \RR$, such that
\begin{equation*}
\rho_i = \frac{T_i\tau}{\tau} \; \; ;
\end{equation*}
therefore equation~(\ref{eq:rho-constr}) defines the second
potential $\tau$:
\begin{equation} \label{eq:tau}
\frac{(T_i T_j\tau)\tau}{(T_i \tau)(T_j\tau)} = 1 -
(T_iQ_{ji})(T_jQ_{ij}) \; , \quad i\ne j \; .
\end{equation}
The potential $\tau$ connecting the forward and backward data is
the $\tau$-{\it function} of the quadrilateral
lattice~\cite{DS-sym}.

Let us find the Laplace transformation of the $\tau$-function.
Formulas \eqref{eq:Lij-Qij} and \eqref{eq:Lij-Qji} imply that
\begin{equation}
1 - (T_i\cL_{ij}(Q_{ji}))(T_j\cL_{ij}(Q_{ij})) =
\frac{Q_{ij}T_iT_jQ_{ij}}{T_jQ_{ij} T_iQ_{ij}} \frac{\tau
T_iT_j\tau}{T_i\tau T_j\tau},
\end{equation}
which, due to equation \eqref{eq:tau}, allows for identification
\begin{equation}
\cL_{ij}(\tau) = \tau Q_{ij}.
\end{equation}
It should be mentioned here that the above formula was strongly
suggested by the identification of the Schlesinger transformation
of the theory of the multicomponent Kadomtsev--Petviashvili
hierarchy~\cite{DKJM,KvL,vL} with the Laplace transformation of
conjugate nets~\cite{DMMMS}.
\begin{Cor}
The geometric meaning of $\tau_{ij}$ as the Laplace transformation
$\cL_{ij}(\tau)$
of the $\tau$-function applies for any dimension of the quadrilateral
lattice.
\end{Cor}

Finally, equation \eqref{eq:tau} rewritten in terms of the
$\tau$-function and its Laplace transformations take the following
form
\begin{equation} \label{eq:Htau}
\tau^{(l)} T_1T_2\tau^{(l)} = (T_1\tau^{(l)})(T_2\tau^{(l)}) -
(T_1 \tau^{(l-1)})(T_2\tau^{(l+1)}) ,
\end{equation}
which is the original Hirota's bilinear form of the discrete Toda
system.

\section{Conclusion}
The geometric interpretation of the Hirota equation (integrable
discrete analog of the Toda system) is given by the Laplace
sequence of quadrilateral lattices, therefore various
representations of the lattice give different versions of the
equation. In the paper we presented four different forms of the
Hirota equation:  (i)~the discrete coupled Volterra
system~\eqref{eq:T3A12}-\eqref{eq:T3A21} for the coefficients of
the Laplace equations, (ii)~the gauge invariant form of the Hirota
equation~\eqref{eq:K} for projective invariants of the Laplace
sequence, (iii)~the discrete Toda system for the rotation
coefficients~\eqref{eq:Q}, and (iv)~the original form of the
Hirota equation~\eqref{eq:Htau} for the $\tau$-function of the
quadrilateral lattice.

\section*{Acknowledgements}

The author would like to thank the organizers of the SIDE III
meeting for invitation and support. 

\providecommand{\bysame}{\leavevmode\hbox to3em{\hrulefill}\thinspace}

\end{document}